\documentclass{article}
\usepackage{graphicx} % Required for inserting images
\usepackage{amsmath} % Math notation
\usepackage{amssymb} % Math notation
\usepackage{authblk} % Affiliation
\usepackage{geometry} % ???
\usepackage{gensymb} % degree symbol

\title{A Global Isometric Embedding of the Reissner-Nordstr\"om Metric into Pseudo-Euclidean Spacetime}

\author[1]{A.T. Eberlein}
\author[1]{C.N. Pope}
\affil[1]{George P. \& Cynthia W. Mitchell Intitute for Fundamental Physics,\linebreak Texas A\&M University, College Station, TX 77843-4242, USA}
\date{}
\newcommand{\RN}{Reissner-Nordstr\"om }
\newcommand{\BL}{Boyer-Lindquist }

\newcommand{\Sch}{Schwarzschild }
\newcommand{\KS}{Kruskal-Szekeres }
\numberwithin{equation}{section}
\begin{document}
\hfill MI-HET-871

{
\let\newpage\relax
\maketitle
}

\begin{abstract}
    The event horizon of the Schwarzschild black hole has been well studied and the singular behavior of the Schwarzschild metric 
    on horizon is understood as a coordinate singularity rather than an essential singularity. One demonstration of this non-singular behavior on horizon was provided by Fronsdal in 1959, by finding a global isometric embedding of the Schwarzschild metric into a six-dimensional pseudo-Euclidean spacetime. Isometric embeddings for the Reissner-Nordstr\"om metric have also been constructed, but they only embed the region external to the inner horizon or in a single Eddington-Finkelstein patch. This paper presents a global isometric embedding for the maximally extended Reissner-Nordstr\"om spacetime into a nine-dimensional pseudo-Euclidean spacetime. We present the solution in terms of explicit local four-dimensional coordinates,  and also as a level-set of functions of the higher-dimensional embedding spacetime. While the Reissner-Nordstr\"om embedding presented has several similarities to the Fronsdal embedding of the Schwarzschild metric, the presence of the second horizon requires additional embedding coordinates and terms not found in the Fronsdal embedding, in order that the embedding is defined and finite on each horizon.
\end{abstract}

\tableofcontents

\section{Introduction}
One way to understand the behavior of a metric near coordinate singularities has been to find a higher-dimensional pseudo-Euclidean spacetime and an embedded submanifold whose inherited metric is the metric of study. For the \Sch metric, this global isometric embedding was found by Fronsdal \cite{Fronsdal}, who also credits Kruskal for the embedding (unpublished). The Fronsdal embedding is well-defined, smooth, finite, and six-dimensional. Six dimensions is known to be optimal, since six dimensions are required for a local embedding, or immersion \cite{Stephani_Kramer_MacCallum_Hoenselaers_Herlt_2003,Eisenhart,JanetorCartan}, so six dimensions is a lower bound for the number of dimensions required for a global embedding.

While the global isometric embedding of the \Sch metric is well understood, the global isometric embedding of the \RN metric has presented more difficulties. Many attempts at obtaining a global isometric embedding of the \RN metric cite the [Cartan-]Janet 
theorem\footnote{\textit{Let }$(M^n,g)$\textit{ be a real-analytic Riemannian manifold,} and $N=\frac{1}{2}n(n+1).$\textit{ Every point of }$M$\textit{ has a neighborhood which has a real-analytic isometric embedding into }$\mathbb{R}^N$\cite{Eisenhart,JanetorCartan}.},
which restricts the dimensions necessary for an isometric immersion to ten. Then, appealing to spherical symmetry \cite{Stephani_Kramer_MacCallum_Hoenselaers_Herlt_2003} reduces the number of necessary dimensions to no more than six. A result from Pandey and Kanel \cite{Pandey_Kansal_1969} requires more than five dimensions. Thus, most efforts to find a global isometric embedding of the \RN metric have been in six dimensions. However, these results are valid only for local isometric embeddings, also known as isometric immersions, so six dimensions serves only as a lower bound for what is required for a global isometric embedding. The works of Friedman \cite{Friedman}, Clarke \cite{Clarke}, and Greene \cite{Greene} give an upper bound on the minimum number of dimensions required for the global isometric embedding of a compact four-dimensional pseudo-Riemannian manifold to $D=48$, 
and to $D=89$ for the non-compact case.

While six dimensions appears to be the lower bound of dimensions needed for a global isometric embedding, attempts at finding a global isometric embedding of the \RN metric in six or more dimensions have failed to give truly global embeddings. The local embeddings by Rosen are either valid only external to the outer horizon \cite{Rosen1} or can instead be made to be valid external to the inner horizon only \cite{Rosen2}. The embedding by Plazowski diverges at both horizons and the embeddings by Paston and Sheykin diverge at the out-going horizons so these fail to be global isometric embeddings \cite{Plazowski,PastonSheykin}. The embedding by Ferraris and Francaviglia \cite{FerrarisFrancaviglia} diverges at the inner horizon, so it also fails to be a global embedding.

It is then of interest to search for a global isometric embedding for the \RN metric where all coordinates remain finite on $r>0,$ for the whole maximal analytic extension, even if a higher number of dimensions is required. In Section \ref{sec: Current State of RN}, the current state of global isometric embeddings for the \RN metric will be reviewed. Section \ref{sec: Definitions} contains some basic conventions and definitions of coordinate charts that will be used in the remainder of the paper.  In Section \ref{sec: Fronsdal} the Fronsdal embedding for the \Sch metric will be provided, highlighting the major features desired for any global isometric embedding. Then in Section \ref{sec: RN Embedding}, a global isometric embedding for the \RN metric will be exhibited with both an explicit embedding in terms of the four coordinates of Reissner-Nordstr\"om, and also an implicit embedding that is expressed independently of the four Reissner-Nordstr\"om coordinates.

\section{The Current State of Global Isometric Embeddings of the \RN Metric} \label{sec: Current State of RN}
In 1965 Rosen \cite{Rosen1} found a six-dimensional embedding where three of the coordinates, slightly adapted, are
\begin{align}
    Z_1^{(-)}&=\omega^{-1}\sqrt{1-\frac{2m}{r}+\frac{q^2}{r^2}}\cos(\omega t)\nonumber\\
    Z_2^{(-)}&=\omega^{-1}\sqrt{1-\frac{2m}{r}+\frac{q^2}{r^2}}\sin(\omega t)\nonumber\\
    Z_3^{(+)}&=\int^r\sqrt{\frac{2mu-q^2+\omega^{-2}\left(\frac{m}{u}-\frac{q^2}{u^2}\right)^2}{u^2-2mu+q^2}}du.
\end{align}
For each of these coordinates, between the inner and outer horizons the square root is imaginary, so this fails to be a global isometric embedding in the region $r_2<r<r_1.$

Later in the same year Rosen \cite{Rosen2} presented a second six-dimensional embedding, where three of the coordinates, slightly adapted, are
\begin{align}
    Z_1^{(-)}&=\kappa^{-1}\sqrt{1-\frac{2m}{r}+\frac{q^2}{r^2}}\sinh(\kappa t)\nonumber\\
    Z_2^{(+)}&=\kappa^{-1}\sqrt{1-\frac{2m}{r}+\frac{q^2}{r^2}}\cosh(\kappa t)\nonumber\\
    Z_3^{(+)}&=\int^r\sqrt{\frac{2mu-q^2-\kappa^{-2}\left(\frac{m}{u}-\frac{q^2}{u^2}\right)^2}{u^2-2mu+q^2}}du.
\end{align}
Between the horizons, the coordinates $Z_1$ and $Z_2$ as presented are imaginary. This problem is removed when correctly giving the coordinates in local coordinates; between the horizons, the coordinates are
\begin{align}
    Z_1^{(-)}&=\kappa^{-1}\sqrt{-1+\frac{2m}{r}-\frac{q^2}{r^2}}\cosh(\kappa t)\nonumber\\
    Z_2^{(+)}&=\kappa^{-1}\sqrt{-1+\frac{2m}{r}-\frac{q^2}{r^2}}\sinh(\kappa t)
\end{align}
The constant $\kappa=r_1^{-1}$ can be chosen to remove the singularity in $Z_3$ at $r=r_1,$ but the singularity at $r_2$ would remain, so this embedding fails to be a global isometric embedding in the region $0<r\le r_2.$ Using Kruskal-Szekeres coordinates in $Z_{1,2}$ would reveal that these coordinates also diverge at the inner horizon due to the behavior as $t\rightarrow\pm\infty.$

Ferraris and Francaviglia in 1980 \cite{FerrarisFrancaviglia} presented a nine-dimensional embedding with five finite coordinates and four dimensions giving an algebraic curve that diverges to the infinity of $\mathbb{R}^4.$ An explicit embedding without any infinite coordinates is not given.

Plazowski in 1972 \cite{Plazowski} gave an eight-dimensional isometric embedding of the \RN metric and extends to the maximal analytic extension with a topological identification. This embedding includes a coordinate
\begin{align}
    Z_8^{(+)}&=\int^r\frac{udu}{u^2-2mu+q^2}\nonumber\\
    &=\frac{r_1}{r_1-r_2}\log\left|\frac{r}{r_1}-1\right|-\frac{r_2}{r_1-r_2}\log\left|\frac{r}{r_2}-1\right|,
\end{align}
which diverges at each horizon.

Paston and Sheykin in 2018 \cite{PastonSheykin} provided three embeddings of the \RN metric which are finite, global, and well-defined for the whole infalling-Eddington-Finkelstein region. These embeddings diverge on the outgoing horizons, so these embedding manifolds fail to have all geodesics. Since the maximal analytic extension of the \RN metric contains all geodesics, it is of interest to find an embedding for the whole maximal analytic extension. The Fronsdal embedding for the Schwarzschild metric is valid for the whole maximal analytic extension of the Schwarzschild manifold.

\section{Definitions} \label{sec: Definitions}
An \textit{embedding into pseudo-Euclidean spacetime} is a map $\mathcal{M}\rightarrow \mathbb{R}^{P,Q}$ where $\mathcal{M}$ is the curved spacetime manifold (typically a four-dimensional manifold). A \textit{coordinate chart} $x^\mu:\mathcal{M}\supseteq U\rightarrow \mathbb{R}^{p,q}$ assigns a set of coordinates to each point on the subset $U$ of the manifold. Typically, the coordinate chart is required to be injective; common exceptions to injectivity are periodic angular coordinates $\mathbb{R}/\mathbb{Z} \cong S^1.$ While many results are easily presented in a coordinate chart, it is uncommon for a manifold to be covered by a single coordinate chart, so multiple presentations should be given in the various coordinate charts that cover the entire manifold.

The \textit{Boyer-Lindquist} coordinate charts use the coordinates $\{t,r,\theta,\phi\}.$ These charts are not defined on the horizon or at the essential singularity, so with two horizons $r_2<r_1,$ there are three \BL charts: $BL_1: r>r_1,$ $BL_2:r_2<r<r_1,$ and $BL_3:0<r<r_2.$

The \textit{Kruskal-Szekeres} coordinate charts use the coordinates $\{V^{KS},U^{KS},\theta,\phi\}.$ These charts are defined on one horizon, but not two simultaneously, so with two horizons there are two \KS charts: $KS_1: r_2<r$ and $KS_2:0<r<r_1.$

An \textit{isometric embedding} is an embedding $\mathcal{M}\rightarrow\mathbb{R}^{P,Q}$ that preserves the metric. A \textit{local} isometric embedding, or an \textit{isometric immersion} is an isometric embedding for a subspace, for example covering one horizon but not the other horizon. A \textit{global} isometric embedding is an isometric embedding given for the entire manifold.

Since it is uncommon to have access to the true points on the manifold $\mathcal{M},$ it is common to give the isometric embedding in terms of the coordinate charts $\mathcal{M}\supseteq U\rightarrow \mathbb{R}^{p,q}\rightarrow\mathbb{R}^{P,Q}.$ Since it is common to require multiple coordinate charts, there will be several presentations of the embedding in the various coordinate charts. These different presentations have to be compatible: if a point $p\in\mathcal{M}$ can be presented in two coordinate charts $x^\mu(p), y^\mu(p),$ the embedded coordinate has to be the same $X^M(x^\mu(p)) = X^M(y^\mu(p)).$

Since the location of horizon(s) for black holes are associated with coordinate singularities, the horizons are part of the manifold and any global isometric embedding has to provide finite embedded coordinates for every horizon simultaneously. Since the essential singularity is not a coordinate singularity, it is not part of the manifold and does not need to have any (finite) embedded coordinate.

\section{Global Isometric Embedding of the Schwarzschild Metric}\label{sec: Fronsdal}
The \Sch metric in \BL coordinates is
\begin{equation}
    ds^2 = -\left(1-\frac{2M}{r}\right)dt^2+\left(1-\frac{2M}{r}\right)^{-1}dr^2+r^2d\theta^2+r^2\sin^2(\theta) d\phi^2.
\end{equation}

The (lightcone) \KS coordinates are defined by
\begin{align}
    t+r+2M\log\left|\frac{r}{2M}-1\right| &= 4M\log\left|\frac{V^{KS}}{2M}\right|\nonumber\\
    t-r-2M\log\left|\frac{r}{2M}-1\right| &= -4M\log\left|\frac{U^{KS}}{2M}\right|
\end{align}

There are four regions (along with their connecting boundaries) that can be covered:
\begin{align}
    I&: r>2M, V^{KS}>0, U^{KS}>0\nonumber\\
    II&: 0<r<2M, V^{KS}>0, U^{KS}<0\nonumber\\
    III&: r>2M, V^{KS}<0, U^{KS}<0\nonumber\\
    IV&: 0<r<2M, V^{KS}<0, U^{KS}>0
\end{align}
In these regions,
\begin{align}
    V^{KS} &= \begin{cases}
    2M\sqrt{\frac{r}{2M}-1}\, e^{(t+r)/4M}&\text{, in }I\\
    2M\sqrt{1-\frac{r}{2M}}\, e^{(t+r)/4M}&\text{, in }II\\
    -2M\sqrt{\frac{r}{2M}-1}\, e^{(t+r)/4M}&\text{, in }III\\
    -2M\sqrt{1-\frac{r}{2M}}\, e^{(t+r)/4M}&\text{, in }IV
    \end{cases}\nonumber\\
    U^{KS} &= \begin{cases}
    2M\sqrt{\frac{r}{2M}-1}\, e^{-(t-r)/4M}&\text{, in }I\\
    -2M\sqrt{1-\frac{r}{2M}}\, e^{-(t-r)/4M}&\text{, in }II\\
    -2M\sqrt{\frac{r}{2M}-1}\, e^{-(t-r)/4M}&\text{, in }III\\
    2M\sqrt{1-\frac{r}{2M}}\, e^{-(t-r)/4M}&\text{, in }IV
    \end{cases}\nonumber\\
    U^{KS}V^{KS} &= 2M(r-2M)e^{r/4M}\text{, everywhere}\nonumber\\
    V^{KS}/U^{KS} &= \begin{cases}
        e^{t/2M}&\text{, in }I,III\\
        -e^{t/2M}&\text{, in }II,IV
    \end{cases}
\end{align}

\begin{figure}
    \centering
    \includegraphics[width=0.7\linewidth]{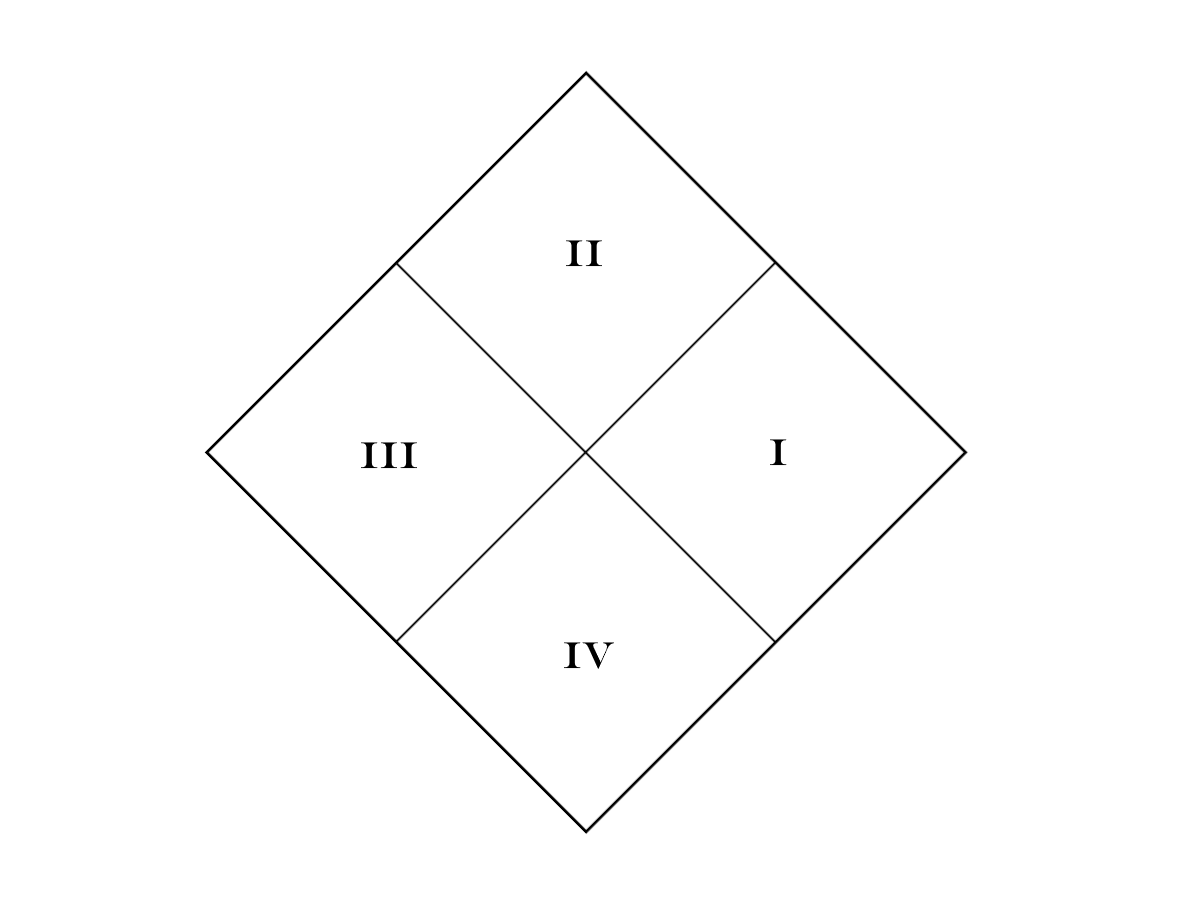}
    \caption{The Kruskal diagram for the maximal analytic extension of the Schwarzschild metric. The $45\degree$ horizon corresponds to $U^{KS}=0$ and the $-45\degree$ horizon corresponds to $V^{KS}=0.$}
    \label{fig:KruskalDiagram_Schwarzschild}
\end{figure}

The \Sch metric in \KS coordinates is then
\begin{equation}
    ds^2 = -\frac{8M}{r}e^{-r/2M}dU^{KS}dV^{KS}+r^2d\theta^2+r^2\sin^2(\theta)d\phi^2,
\end{equation}
where $r=r(V,U)$ is implicitly given in terms of $U$ and $V$ by the definitions.

The Fronsdal embedding \cite{Fronsdal} is a global isometric embedding of this maximal analytic extension of the \Sch metric into six-dimensional Minkowski spacetime. To properly give the embedding, it should be presented in the \KS chart, but it is also typically given the the two \BL charts: $BL_+: r>2M,$ and $BL_-: 0<r<2M.$ The Fronsdal embedding is then
\begin{align}
    X^{(+)}&=r\sin\theta\cos\phi\nonumber\\
    Y^{(+)}&=r\sin\theta\sin\phi\nonumber\\
    Z^{(+)}&=r\cos\theta\nonumber\\
    R^{(+)}&=\int^r\sqrt{\frac{2M}{u}+\frac{4M^2}{u^2}+\frac{8M^3}{u^3}}du.
\end{align}
The last two coordinates have different presentations in different regions and in different charts
\begin{align}
    T^{(-)}&=\begin{cases}
        4M\sqrt{1-\frac{2M}{r}}\sinh(t/4M)&\text{, in }I(BL_+)\\
        4M\sqrt{\frac{2M}{r}-1}\cosh(t/4M)&\text{, in }II(BL_-)\\
        -4M\sqrt{1-\frac{2M}{r}}\sinh(t/4M)&\text{, in }III(BL_+)\\
        -4M\sqrt{\frac{2M}{r}-1}\cosh(t/4M)&\text{, in }IV(BL_-)\\
        \sqrt{\frac{2M}{r}}e^{-r/4M}(V^{KS}-U^{KS})&\text{, everywhere }(KS)\\
    \end{cases}\nonumber\\
    S^{(+)}&=\begin{cases}
        4M\sqrt{1-\frac{2M}{r}}\cosh(t/4M)&\text{, in }I(BL_+)\\
        4M\sqrt{\frac{2M}{r}-1}\sinh(t/4M)&\text{, in }II(BL_-)\\
        -4M\sqrt{1-\frac{2M}{r}}\cosh(t/4M)&\text{, in }III(BL_+)\\
        -4M\sqrt{\frac{2M}{r}-1}\sinh(t/4M)&\text{, in }IV(BL_-)\\
        \sqrt{\frac{2M}{r}}e^{-r/4M}(V^{KS}+U^{KS})&\text{, everywhere }(KS)\\
    \end{cases}.
\end{align}

This gives a global isometric embedding for the maximal analytic extension of the \Sch metric. A global isometric embedding for the simple \Sch metric keeps only the regions $I$ and $II.$

In Fronsdal's original paper \cite{Fronsdal}, this embedding was also given in the four-coordinate-free manner,
\begin{align}
    X^2+Y^2+Z^2&=r^2\nonumber\\
    S^2-T^2&=16M^2\left(1-\frac{2M}{r}\right)\nonumber\\
    R(r) &= \int^r\sqrt{\frac{2M}{u}+\frac{4M^2}{u^2}+\frac{8M^3}{u^3}}du,
\end{align}
where since $R(r)$ is injective, it can be inverted to get $r(R).$

Allowing for additional coordinates, the metric can be written
\begin{equation}
    ds^2=dX^2+dY^2+dZ^2-dT^2+dS^2+dR_1^2+dR_2^2+dR_3^2
\end{equation}
where
\begin{align}
    R_1&=\int^r\sqrt{\frac{2M}{u}}du=2\sqrt{2Mr}\nonumber\\
    R_2&=\int^r\frac{2M}{u}du=2M\log\left(\frac{r}{2M}\right)\nonumber\\
    R_3&=-\int^r\left(\frac{2M}{u}\right)^{3/2}du=4M\sqrt{\frac{2M}{r}}.
\end{align}

So a four-coordinate-free presentation of the embedding can be given by
\begin{align}
    R_1&=4Me^{R_2/4M}\nonumber\\
    R_3&=4Me^{-R_2/4M}\nonumber\\
    X^2+Y^2+Z^2&=4M^2e^{R_2/M}\nonumber\\
    S^2-T^2&=16M^2\left(1-e^{-R_2/2M}\right).
\end{align}

This four-coordinate-free presentation of the embedding gives the analogue of the coordinate-free presentation of the spherical metric as the induced metric on $X^2+Y^2+Z^2=r^2.$

If we define light-cone coordinates $V=T+S, U=T-S,$ then we have $-dT^2+dS^2=-dVdU,$ and the new light-cone pseudo-Euclidean coordinates $\{V,U\}$ are
\begin{align}
    V&=\begin{cases}
        4M\sqrt{1-\frac{2M}{r}}e^{t/4M}&\text{, in }I(BL_+)\\
        4M\sqrt{\frac{2M}{r}-1}e^{t/4M}&\text{, in }II(BL_-)\\
        -4M\sqrt{1-\frac{2M}{r}}e^{t/4M}&\text{, in }III(BL_+)\\
        -4M\sqrt{\frac{2M}{r}}e^{t/4M}&\text{, in }IV(BL_-)\\
        2\sqrt{\frac{2M}{r}}e^{-r/4M}V^{KS}&\text{, everywhere }(KS)
    \end{cases}\nonumber\\
    U&=\begin{cases}
        -4M\sqrt{1-\frac{2M}{r}}e^{-t/4M}&\text{, in }I(BL_+)\\
        4M\sqrt{\frac{2M}{r}-1}e^{-t/4M}&\text{, in }II(BL_-)\\
        4M\sqrt{1-\frac{2M}{r}}e^{-t/4M}&\text{, in }III(BL_+)\\
        -4M\sqrt{\frac{2M}{r}}e^{-t/4M}&\text{, in }IV(BL_-)\\
        -2\sqrt{\frac{2M}{r}}e^{-r/4M}U^{KS}&\text{, everywhere }(KS).
    \end{cases}
\end{align}

These coordinates satisfy the four-coordinate-free presentation
\begin{align}
    R_1&=4Me^{R_2/4M}\nonumber\\
    R_3&=4Me^{-R_2/4M}\nonumber\\
    X^2+Y^2+Z^2&=4M^2e^{R_2/M}\nonumber\\
    -UV&=16M^2\left(1-e^{-R_2/2M}\right).
\end{align}

\section{Global Isometric Embedding of the Reissner-Nordstr\"om Metric}\label{sec: RN Embedding}
The \RN metric in \BL coordinates is
\begin{equation}
    ds^2=-\left(1-\frac{2M}{r}+\frac{Q^2}{r^2}\right)dt^2+\left(1-\frac{2M}{r}+\frac{Q^2}{r^2}\right)^{-1}+r^2d\theta^2+r^2\sin^2\theta d\phi^2.
\end{equation}
Here, $Q^2=q^2+p^2,$ where $q$ is the parameter associated with electric charge and $p$ is the parameter associated with magnetic charge. The locations of the horizons are $r_{1,2}=M\pm\sqrt{M^2-Q^2}$ with $r_1>r_2>0.$ The surface gravities at each horizon are $\kappa_i = \frac{r_1-r_2}{2r_i^2},$ and the horizon residues are $R_i = \frac{1}{2\kappa_i}=\frac{r_i^2}{r_1-r_2}.$ That is, $g_{rr}=1+\frac{R_1}{r-r_1}-\frac{R_2}{r-r_2}.$

Since the \BL coordinate $t$ diverges at each horizon, local coordinates defined at each horizon are required. The local (lightcone) \KS coordinates are $\{V_i^{KS},U_i^{KS}\}$ defined by
\begin{align}
    t+r+R_1\log\left|\frac{r}{r_1}-1\right|-R_2\log\left|\frac{r}{r_2}-1\right| &= 2R_1\log\left|\frac{V_1^{KS}}{r_1}\right|\nonumber\\
    &=-2R_2\log\left|\frac{U_2^{KS}}{r_2}\right|\nonumber\\
    t-r-R_1\log\left|\frac{r}{r_1}-1\right|+R_2\log\left|\frac{r}{r_2}-1\right| &= -2R_1\log\left|\frac{U_1^{KS}}{r_1}\right|\nonumber\\
    &=2R_2\log\left|\frac{V_2^{KS}}{r_2}\right|.
\end{align}

For the maximal analytic extension of the \RN metric, there are infinitely many regions to cover, coming in three types: $A_m,$ $B_m,$ and $C_m.$ In these regions, there are different sets of local \BL coordinates: $BL_1,$ $BL_2,$ and $ BL_3.$

\begin{figure}
    \centering
    \includegraphics[width=0.25\linewidth]{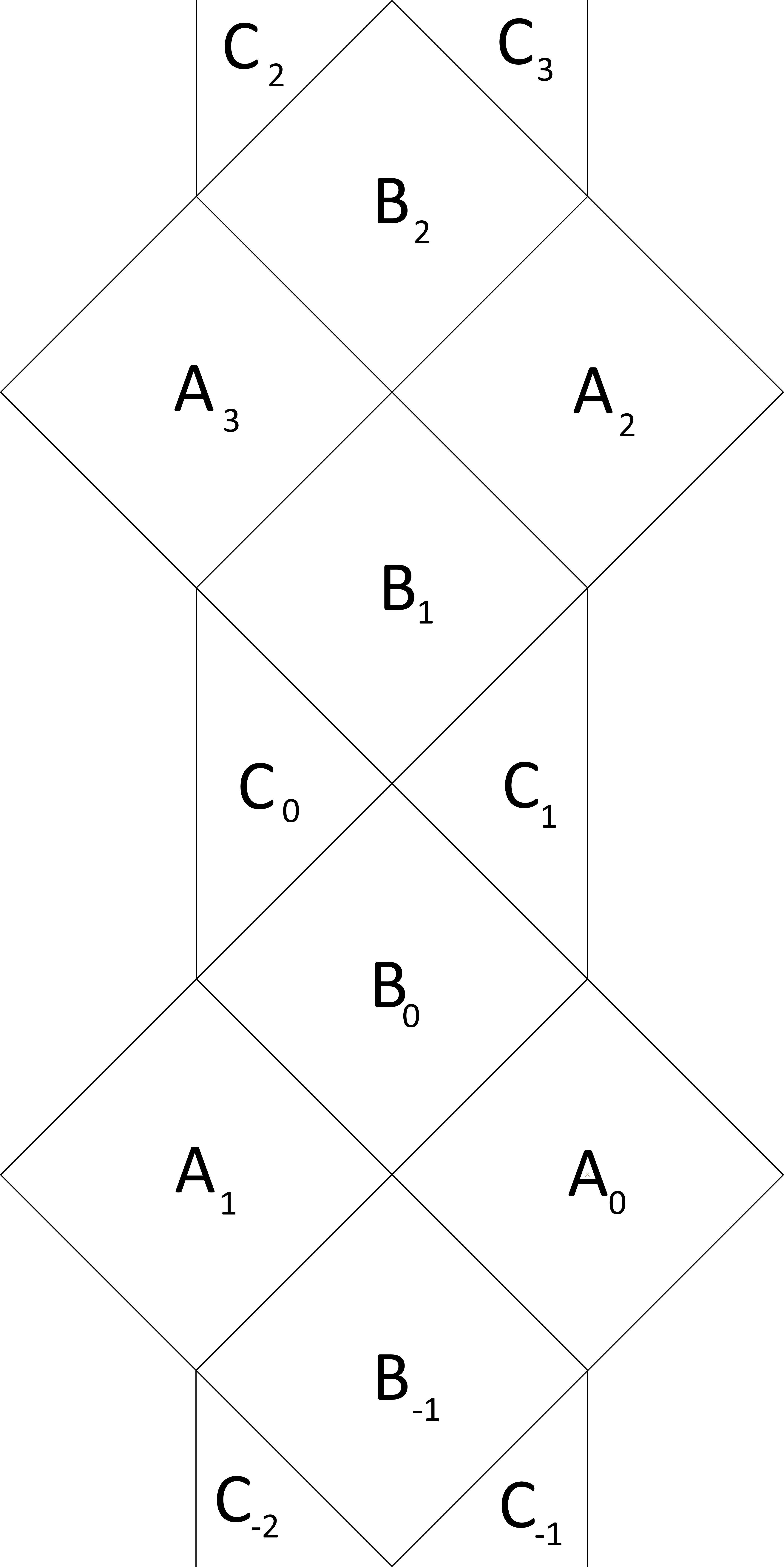}
    \caption{The Kruskal diagram for the maximally extended \RN manifold.}
    \label{fig:RNKruskal}
\end{figure}

A global isometric embedding into pseudo-Euclidean spacetime is given by
\begin{equation}
    ds^2=dX^2+dY^2+dZ^2+dV_1dU_1-dV_2dU_2+dR_+^2-dR_-^2
\end{equation}
where
\begin{align}\label{eqn: RN_embed_XYZ}
    X^{(+)}&=r\sin\theta\cos\phi\nonumber\\
    Y^{(+)}&=r\sin\theta\sin\phi\nonumber\\
    Z^{(+)}&=r\cos\theta.
\end{align}

The embedding coordinates $V_{1,2},U_{1,2}$ in the local \BL coordinates are, in $A_{2m+\epsilon},$
\begin{align}\label{eqn: RN_embed_UV_A}
    V_1&=e^{\pi i(m-\epsilon+2m\kappa_1R_2)}\Omega^{-1/2}\frac{2R_1}{r}\sqrt{r-r_1}(r-r_2)^pe^{\kappa_1t}\nonumber\\
    U_1&=e^{\pi i(m-\epsilon-2m\kappa_1R_2)}\Omega^{-1/2}\frac{2R_1}{r}\sqrt{r-r_1}(r-r_2)^pe^{-\kappa_1t}\nonumber\\
    V_2&=e^{\pi i(m+2(m-\epsilon)\kappa_2R_1)}\Omega^{-1/2}\frac{2R_2}{r}(r-r_1)^p\sqrt{r-r_2}e^{\kappa_2t}\nonumber\\
    U_2&=e^{\pi i(m-2(m-\epsilon)\kappa_2R_1)}\Omega^{-1/2}\frac{2R_2}{r}(r-r_1)^p\sqrt{r-r_2}e^{-\kappa_2t}.
\end{align}

In $B_{2m+\epsilon},$
\begin{align}\label{eqn: RN_embed_UV_B}
    V_1&=e^{\pi i(m+2(m+\epsilon)\kappa_1R_2)}\Omega^{-1/2}\frac{2R_1}{r}\sqrt{r_1-r}(r-r_2)^pe^{\kappa_1t}\nonumber\\
    U_1&=-e^{\pi i(m-2(m+\epsilon)\kappa_1R_2)}\Omega^{-1/2}\frac{2R_1}{r}\sqrt{r_1-r}(r-r_2)^pe^{-\kappa_1t}\nonumber\\
    V_2&=e^{\pi i(p+m+\epsilon+(2m-1)\kappa_2R_1)}\Omega^{-1/2}\frac{2R_2}{r}(r_1-r)^p\sqrt{r-r_2}e^{\kappa_2t}\nonumber\\
    U_2&=e^{\pi i(p+m+\epsilon-(2m-1)\kappa_2R_1)}\Omega^{-1/2}\frac{2R_2}{r}(r_1-r)^p\sqrt{r-r_2}e^{-\kappa_2t}.
\end{align}

In $C_{2m+\epsilon},$
\begin{align}\label{eqn: RN_embed_UV_C}
    V_1&=e^{\pi i(p+m+[2(m+\epsilon)-1]\kappa_1R_2)}\Omega^{-1/2}\frac{2R_1}{r}\sqrt{r_1-r}(r_2-r)^pe^{\kappa_1t}\nonumber\\
    U_1&=-e^{\pi i(p+m-[2(m+\epsilon)-1]\kappa_1R_2)}\Omega^{-1/2}\frac{2R_1}{r}\sqrt{r_1-r}(r_2-r)^pe^{-\kappa_1t}\nonumber\\
    V_2&=e^{\pi i(p+m+\epsilon+(2m-1)\kappa_2R_1)}\Omega^{-1/2}\frac{2R_2}{r}(r_1-r)^p\sqrt{r_2-r}e^{\kappa_2t}\nonumber\\
    U_2&=-e^{\pi i(p+m+\epsilon-(2m-1)\kappa_2R_1)}\Omega^{-1/2}\frac{2R_2}{r}(r_1-r)^p\sqrt{r_2-r}e^{-\kappa_2t},
\end{align}
where
\begin{align}
    \Omega&=(r-r_2)^{2p-1}-(r-r_1)^{2p-1}>0\\
    \mathbb{Z}_+\ni p&\ge \max(k,\kappa_1R_2,\kappa_2R_1)=\max(k,r_1^2/(2r_2^2))\nonumber
\end{align}
where $V_i, U_i$ are of class $C^k.$

These coordinates satisfy
\begin{equation}
    \kappa_1^{-2}V_1U_1-\kappa_2^{-2}V_2U_2=1-\frac{r_1+r_2}{r}+\frac{r_1r_2}{r^2}=1-\frac{2M}{r}+\frac{Q^2}{r^2}.
\end{equation}

The complex exponential coefficients in front of each term have been chosen so the coordinates are smoothly defined when crossing each horizon, and also serves to separate each region in the maximally extended spacetime. The presentation of these embedding coordinates in local \KS coordinates is found in Appendix A.

The last two coordinates $R_\pm$ satisfy
\begin{multline}
    (R'_+)^2-(R'_-)^2 = \left\{\frac{R_1}{r-r_1}-\left[\left(\Omega^{-1/2}\frac{2R_1}{r}\sqrt{r-r_1}(r-r_2)^p\right)'\right]^2\right\}\\
    -\left\{\frac{R_2}{r-r_2}-\left[\left(\Omega^{-1/2}\frac{2R_2}{r}(r-r_1)^p\sqrt{r-r_2}\right)'\right]^2\right\}.
\end{multline}

By construction, this is a rational function with no poles, so it can be separated into a strictly positive part and a strictly negative part. If
\begin{equation}
    (R'_+)^2-(R'_-)^2 = Q_+(r)-Q_-(r)
\end{equation}
with $Q_\pm(r)>0,$ then 
\begin{align}\label{eqn: RN_embed_Rpm}
    R_+&=\int^r\sqrt{Q_+(u)}du\nonumber\\
    R_-&=\int^r\sqrt{Q_-(u)}du.
\end{align}

For example, for $p=1,$
\begin{align}
    (R_+')^2-(R_-')^2&=\frac{-2r_1r_2(r_1+r_2)r^3-r_1r_2(4r_1^2+3r\nonumber_1r_2+4r_2^2)+4r_1^2r_2^2(r_1^2+r_1r_2+r_2^2)}{(r_1-r_2)^2r^4}\nonumber\\
    Q_+(r)&=\frac{4r_1^2r_2^2(r_1^3-r_2^3)}{(r_1-r_2)^3r^4}\nonumber\\
    Q_-(r)&=\frac{r_1r_2\left[2(r_1+r_2)r+(4r_1^2+3r_1r_2+4r_2^2)\right]}{(r_1-r_2)^2r^2}\nonumber\\
    R_+&=\sqrt{\frac{r_1^3-r_2^3}{(r_1-r_2)^3}}\frac{2r_1r_2}{r}\nonumber\\
    R_-&=\int^r\sqrt{\frac{r_1r_2\left[2(r_1+r_2)u+(4r_1^2+3r_1r_2+4r_2^2)\right]}{(r_1-r_2)^2u^2}}du.
\end{align}

This gives an embedding of the \RN metric into $\mathbb{R}^{1,5}\times\mathbb{C}^4.$ By allowing complex coordinates, so long as $r_2^2/r_1^2\notin \mathbb{Q},$ this is a global isometric embedding of the entire maximal analytic extension of the \RN metric.

Since the integrands in $R_\pm$ are non-negative, $R_\pm$ are injective so one can be inverted to give $r(R_+).$ We then have four-coordinate-free embedding equations for the \RN spacetime:
\begin{align}\label{eqn: RN_embed_levelset}
    R_-(r)&=R_-(r(R_+))\nonumber\\
    X^2+Y^2+Z^2&=r^2\nonumber\\
    V_1U_1&=\Omega^{-1}\frac{4R_1^2}{r^2}(r-r_1)(r-r_2)^{2p}\nonumber\\
    V_2U_2&=\Omega^{-1}\frac{4R_2^2}{r^2}(r-r_1)^{2p}(r-r_2)\nonumber\\
    \left[\Omega^{1/2}\left(\frac{r}{2R_1}\right)(r-r_1)^{-1/2}(r-r_2)^{-p}V_1\right]&=\left[\Omega^{1/2}\left(\frac{r}{2R_2}\right)(r-r_1)^{-p}(r-r_2)^{-1/2}V_2\right]^{2\kappa_1R_2}
\end{align}
with $r=r(R_+).$ This last equation is symmetric in $r_1\leftrightarrow r_2$; if $\kappa$ is arbitrary
\begin{equation}
    \left[\Omega^{1/2}\left(\frac{r}{2R_1}\right)(r-r_1)^{-1/2}(r-r_2)^{-p}V_1\right]^{\kappa_2/\kappa}=\left[\Omega^{1/2}\left(\frac{r}{2R_2}\right)(r-r_1)^{-p}(r-r_2)^{-1/2}V_2\right]^{\kappa_1/\kappa}.
\end{equation}

If $D_m$ is any of $A_m,B_m,C_m,$ there is a simple relation between $D_m$ and $D_{m+4}:$
\begin{align}
    V_i(D_{m+4})&=V_i(D_m)e^{4\pi i\kappa_\pm R_\mp}\nonumber\\
    U_i(D_{m+4})&=U_i(D_m)e^{-4\pi i\kappa_\pm R_\mp}.
\end{align}

Identification of $D_{m+4}\sim D_m$ corresponds to identifying
\begin{equation}
    e^{\kappa_it}\sim e^{\kappa_i(t-2\pi i\kappa_j^{-1})}.
\end{equation}

The choice of the surface gravity in $e^{\kappa_1t},$ for example, was required for this embedding to be defined on each horizon in local \KS coordinates while also giving the simple pole at each horizon in $g_{rr}.$ This choice together with the above identification gives the correct periodicity required to remove the conical singularity near the corresponding horizon in the metric after a Wick rotation of time.

\section{Discussion and Conclusions}\label{sec: Discussion}

A global isometric embedding for the Reissner-Nordstr\"om metric has been demonstrated. An embedding given in terms of local Boyer-Lindquist coordinates was provided by Equation \ref{eqn: RN_embed_XYZ} through Equation \ref{eqn: RN_embed_UV_C} and Equation \ref{eqn: RN_embed_Rpm}. This embedding given in terms of local \KS coordinates is provided in Appendix \ref{app: KS coordinate embedding}. The embedding was also given as a level set of five functions in Equation \ref{eqn: RN_embed_levelset}. This embedding mirrors the Fronsdal embedding for the \Sch metric in several ways, but the presence of the second horizon increases the number of dimensions of this embedding. Special care was needed for the embedding coordinates to be properly defined on each horizon. The global isometric embedding presented here is an embedding into nine-dimensional pseudo-Euclidean spacetime, but an embedding into a lower dimensional pseudo-Euclidean spacetime may be possible.

The approach in this paper does not work for the extremal \RN metric where the inner and outer horizon coincide. Further research is necessary for the global isometric embedding of non-simple poles of $g_{rr}.$ A global isometric embedding provides a different avenue of study for metrics, exchanging the differential geometry of curved spacetimes with pseudo-Euclidean geometry restricted on a submanifold, and the study of properties of a curved spacetime can be replaced with the study of properties of the embedded submanifold. Therefore, it may be of interest to find global isometric embeddings of other interesting metrics. The approach in this paper may be applicable to other two-horizon metrics (e.g. Schwarzschild-de Sitter, Reissner-Nordstr\"om-anti-de Sitter), metrics with more than two horizons (e.g. Reissner-Nordstr\"om-de Sitter), and higher-dimensional analogues. It may also be of interest to find global isometric embeddings for axisymmetric metrics (e.g. Kerr, Kerr-Newman) and metrics that depend on time (e.g.  FLRW).

% \section{Discussion}\label{sec: Discussion}

% A global isometric embedding for the \RN metric has been demonstrated in nine dimensions. The embedding was presented in two different ways: an explicit embedding in the local four-coordinates that was complex valued, and an implicit embedding given by equations of the embedding coordinates. Each embedding was well-behaved on each horizon, showing that the horizons are coordinate singularities rather than essential singularities.

% This paper showed a global isometric embedding for both the \Sch metric and the \RN metric. It remains to be done to find a global isometric embedding for other known solutions, such as adding a cosmological constant or a rotation parameter or adding additional fields changing the energy-momentum tensor.

\section*{Acknowledgements}\label{sec: Acknowledgements}
This work was supported in part by DOE grant DE-SC0010813.

\appendix
\section{Embedding Coordinates in Local \KS Coordinates}\label{app: KS coordinate embedding}

Local \KS coordinates can be defined as
\begin{align}
    V_1^{KS}&=\begin{cases}
        (-1)^m r_1\sqrt{\frac{r}{r_1}-1}\left(\frac{r}{r_2}-1\right)^{-\kappa_1R_2}e^{\kappa_1(t+r)}&\text{, in }A_m\\
        (-1)^m r_1\sqrt{1-\frac{r}{r_1}}\left(\frac{r}{r_2}-1\right)^{-\kappa_1R_2}e^{\kappa_1(t+r)}&\text{, in }B_m
    \end{cases}\nonumber\\
    U_1^{KS}&=\begin{cases}
        (-1)^m r_1\sqrt{\frac{r}{r_1}-1}\left(\frac{r}{r_2}-1\right)^{-\kappa_1R_2}e^{-\kappa_1(t-r)}&\text{, in }A_m\\
        -(-1)^m r_1\sqrt{1-\frac{r}{r_1}}\left(\frac{r}{r_2}-1\right)^{-\kappa_1R_2}e^{-\kappa_1(t-r)}&\text{, in }B_m
    \end{cases}\nonumber\\
    V_2^{KS}&=\begin{cases}
        (-1)^mr_2\left(1-\frac{r}{r_1}\right)^{-\kappa_2R_1}\sqrt{\frac{r}{r_2}-1}e^{\kappa_2(t+r)}&\text{, in }B_m\\
        (-1)^mr_2\left(1-\frac{r}{r_1}\right)^{-\kappa_2R_1}\sqrt{1-\frac{r}{r_2}}e^{\kappa_2(t+r)}&\text{, in }C_m
    \end{cases}\nonumber\\
    U_2^{KS}&=\begin{cases}
        (-1)^mr_2\left(1-\frac{r}{r_1}\right)^{-\kappa_2R_1}\sqrt{\frac{r}{r_2}-1}e^{-\kappa_2(t-r)}&\text{, in }B_m\\
        -(-1)^mr_2\left(1-\frac{r}{r_1}\right)^{-\kappa_2R_1}\sqrt{1-\frac{r}{r_2}}e^{-\kappa_2(t-r)}&\text{, in }C_m
    \end{cases}.
\end{align}

The global isometric embedding must also be given in local \KS coordinates to be defined on each horizon. In $A_{2m+\epsilon},$ in $KS_1,$
\begin{align}
    V_1&=e^{\pi i(m+2m\kappa_1R_2)}\frac{2R_1r_1^{-1/2}r_2^p}{r\sqrt{\Omega}}\left(\frac{r}{r_2}-1\right)^{p+\kappa_1R_2} e^{-\kappa_1r}V_1^{KS}\nonumber\\
    U_1&=e^{\pi i(m-2m\kappa_1R_2)}\frac{2R_1r_1^{-1/2}r_2^p}{r\sqrt{\Omega}}\left(\frac{r}{r_2}-1\right)^{p+\kappa_1R_2}e^{-\kappa_1r}U_1^{KS}\nonumber\\
    V_2&=e^{\pi i(m+2(m-\epsilon)\kappa_2R_1)}\frac{2R_2r_1^pr_2^{1/2}}{r\sqrt{\Omega}}\left(\frac{r}{r_1}-1\right)^{p-\kappa_2R_1}\left(\frac{r}{r_2}-1\right)e^{-\kappa_2r}\left(e^{\pi i\epsilon}\frac{V_1^{KS}}{r_1}\right)^{2\kappa_2R_1}\nonumber\\
    U_2&=e^{\pi i(m-2(m-\epsilon)\kappa_2R_1)}\frac{2R_2r_1^pr_2^{1/2}}{r\sqrt{\Omega}}\left(\frac{r}{r_1}-1\right)^{p-\kappa_2R_1}\left(\frac{r}{r_2}-1\right)e^{-\kappa_2r}\left(e^{-\pi i\epsilon}\frac{U_1^{KS}}{r_1}\right)^{2\kappa_2R_1}.
\end{align}

In $B_{2m+\epsilon},$ in $KS_1,$
\begin{align}
    V_1&=e^{\pi i(m+\epsilon+2(m+\epsilon)\kappa_1R_2)}\frac{2R_1r_1^{-1/2}r_2^p}{r\sqrt{\Omega}}\left(\frac{r}{r_2}-1\right)^{p+\kappa_1R_2} e^{-\kappa_1r}V_1^{KS}\nonumber\\
    U_1&=e^{\pi i(m+\epsilon-2(m+\epsilon)\kappa_1R_2)}\frac{2R_1r_1^{-1/2}r_2^p}{r\sqrt{\Omega}}\left(\frac{r}{r_2}-1\right)^{p+\kappa_1R_2}e^{-\kappa_1r}U_1^{KS}\nonumber\\
    V_2&=e^{\pi i(p+m+\epsilon+(2m-1)\kappa_2R_1)}\frac{2R_2r_1^pr_2^{1/2}}{r\sqrt{\Omega}}\left(1-\frac{r}{r_1}\right)^{p-\kappa_2R_1}\left(\frac{r}{r_2}-1\right)e^{-\kappa_2r}\left(e^{\pi i\epsilon}\frac{V_1^{KS}}{r_1}\right)^{2\kappa_2R_1}\nonumber\\
    U_2&=e^{\pi i(p+m+\epsilon-(2m-1)\kappa_2R_1)}\frac{2R_2r_1^pr_2^{1/2}}{r\sqrt{\Omega}}\left(1-\frac{r}{r_1}\right)^{p-\kappa_2R_1}\left(\frac{r}{r_2}-1\right)e^{-\kappa_2r}\left(e^{-\pi i(\epsilon+1)}\frac{U_1^{KS}}{r_1}\right)^{2\kappa_2R_1}.
\end{align}

In $B_{2m+\epsilon},$ in $KS_2,$
\begin{align}
    V_1&=e^{\pi i(m+2(m+\epsilon)\kappa_1R_2)}\frac{2R_1r_1^{1/2}r_2^p}{r\sqrt{\Omega}}\left(1-\frac{r}{r_1}\right)\left(\frac{r}{r_2}-1\right)^{p-\kappa_1R_2}e^{-\kappa_1r}\left(e^{-\pi i\epsilon}\frac{V_2^{KS}}{r_2}\right)^{2\kappa_1R_2}\nonumber\\
    U_1&=-e^{\pi i(m-2(m+\epsilon)\kappa_1R_2)}\frac{2R_1r_1^pr_2^{1/2}}{r\sqrt{\Omega}}\left(1-\frac{r}{r_1}\right)\left(\frac{r}{r_2}-1\right)^{p-\kappa_1R_2}e^{-\kappa_1r}\left(e^{-\pi i\epsilon}\frac{U_2^{KS}}{r_2}\right)^{2\kappa_1R_2}\nonumber\\
    V_2&=e^{\pi i(p+m+(2m-1)\kappa_2R_1)}\frac{2R_2r_1^pr_2^{-1/2}}{r\sqrt{\Omega}}\left(1-\frac{r}{r_1}\right)^{p+\kappa_2R_1}e^{-\kappa_2r}V_2^{KS}\nonumber\\
    U_2&=e^{\pi i(p+m-(2m-1)\kappa_2R_1)}\frac{2R_2r_1^pr_2^{-1/2}}{r\sqrt{\Omega}}\left(1-\frac{r}{r_1}\right)^{p+\kappa_2R_1}e^{-\kappa_2r}U_2^{KS}.
\end{align}

In $C_{2m+\epsilon},$ in $KS_2,$
\begin{align}
    V_1&=e^{\pi i(p+m+[2(m+\epsilon)-1]\kappa_1R_2)}\frac{2R_1r_1^{1/2}r_2^p}{r\sqrt{\Omega}}\left(1-\frac{r}{r_1}\right)\left(1-\frac{r}{r_2}\right)^{p-\kappa_1R_2}e^{-\kappa_1r}\left(e^{-\pi i\epsilon}\frac{V_2^{KS}}{r_2}\right)^{2\kappa_1R_2}\nonumber\\
    U_1&=-e^{\pi i(p+m-[2(m+\epsilon)-1]\kappa_1R_2)}\frac{2R_1r_1^{1/2}r_2^p}{r\sqrt{\Omega}}\left(1-\frac{r}{r_1}\right)\left(1-\frac{r}{r_2}\right)^{p-\kappa_1R_2}e^{-\kappa_1r}\left(e^{-\pi i(\epsilon+1)}\frac{U_2^{KS}}{r_2}\right)^{2\kappa_1R_2}\nonumber\\
    V_2&=e^{\pi i(p+m+(2m-1)\kappa_2R_1)}\frac{2R_2r_1^pr_2^{-1/2}}{r\sqrt{\Omega}}\left(1-\frac{r}{r_1}\right)^{p+\kappa_2R_1}e^{-\kappa_2r}V_2^{KS}\nonumber\\
    U_2&=e^{\pi i(p+m-(2m-1)\kappa_2R_1)}\frac{2R_2r_1^pr_2^{-1/2}}{r\sqrt{\Omega}}\left(1-\frac{r}{r_1}\right)^{p+\kappa_2R_1}e^{-\kappa_2r}U_2^{KS}.
\end{align}

\newpage

\bibliography{biblio}
\bibliographystyle{alpha}
\end{document}